\documentstyle[prl,aps,multicol,psfig]{revtex}

\def\be{\begin{equation}}
\def\ee{\end{equation}}
\def\br{\begin{eqnarray}}
\def\er{\end{eqnarray}}
\def\NPB#1#2#3{{\it Nucl.~Phys.} {\bf{B#1}} (19#2) #3}
\def\PLB#1#2#3{{\it Phys.~Lett.} {\bf{B#1}} (19#2) #3}
\def\PRD#1#2#3{{\it Phys.~Rev.} {\bf{D#1}} (19#2) #3}
\def\PRL#1#2#3{{\it Phys.~Rev.~Lett.} {\bf{#1}} (19#2) #3}

\def\PRD#1#2#3{{\it Phys.~Rev.} {\bf{D#1}} (19#2) #3}
\def\PRL#1#2#3{{\it Phys.~Rev.~Lett.} {\bf{#1}} (19#2) #3}

\renewcommand{\narrowtext}{\begin{multicols}{2} \global\columnwidth20.5pc}
\renewcommand{\widetext}{\end{multicols} \global\columnwidth42.5pc}
\begin{document}

\title{Domain walls in supersymmetric QCD: from weak to strong coupling}
\author{B. de Carlos\footnote{e-mail address: Beatriz.de.Carlos@cern.ch}}
\address{Theory Division, CERN, CH-1211 Geneva 23, Switzerland}
\author{J.M. Moreno\footnote{e-mail address: jesus@makoki.iem.csic.es}}
\address{Instituto de Estructura de la Materia, CSIC, Serrano 123, 28006 
Madrid, Spain}

\maketitle

\begin{abstract}

We consider domain walls that appear in supersymmetric QCD with 
$N_f < N_c$ massive flavours. In particular, for $2 N_f < N_c$ we 
explicitly construct the domain walls that interpolate between vacua 
labeled by $i$ and $i+ N_f$. We show that these solutions are 
Bogomol'nyi-Prasad-Sommerfield (BPS) saturated for any value of the 
mass of the matter fields. This fact allows us to evaluate the 
large mass limit of these domain walls. We comment on the relevance of 
these solutions for supersymmetric gluodynamics.

\end{abstract}

\vspace{0.2cm} 

PACS numbers: 11.30.Pb, 11.27.+d \hspace{3.2cm} Preprint numbers:
CERN-TH/99-150, IEM-FT-194/99


\narrowtext

In recent times, a lot of attention has been drawn to the existence of exact
solutions in supersymmetric gauge theories that are in the strong coupling 
regime. 
One of the more relevant issues is that of domain walls in SU(N) 
supersymmetric gluodynamics, the theory of gluons and gluinos. Those arise 
because this theory has an axial U(1) symmetry broken by the anomaly to a 
discrete $Z_{2N}$ chiral symmetry. Due to non-perturbative effects gluino 
condensates $\left( \langle \lambda \lambda \rangle \right)$ form, breaking 
the symmetry further down to $Z_2$. This leaves us with a set of $N$ 
different vacua labelled by
\be
\langle {\rm Tr} \lambda \lambda \rangle = \Lambda^3 e^{2 \pi i k/N} \;\; 
k= 0,1,...,N-1 \;,
\ee
where $\Lambda$ is the condensation scale, and, as indicated above, a set
of domain walls interpolating between 
them (in Ref.~\cite{KovnerS0} it was pointed out the existence of
a chirally symmetric vacuum where the gaugino condensate vanishes. 
Here we will only consider domain walls involving chirally asymmetric 
vacua). If we asume that they are BPS saturated, the energy density of 
these walls is exactly calculable and given by 
\cite{Dvali97,Kovner97,Chibisov98,Shifm98}
\be
\epsilon = \frac{N}{8 \pi^2} | \langle {\rm Tr} \lambda \lambda 
\rangle_{\infty} - \langle {\rm Tr} \lambda \lambda \rangle_{-\infty}| \;\;,
\ee
In fact, it has been suggested in Ref.~\cite{Dvali99} that, in the large 
$N$ limit, these domain walls are BPS states. On the other hand, these 
solutions preserving half of the supersymmetry would play an important role 
in the D-brane description of $N=1$ SQCD~\cite{Witten97}. However, we want 
to stress that whether or not these configurations were BPS saturated was,
up to now, still an open question \cite{Shifm98}. 

A useful way of gaining intuition on pure gluodynamics is by adding 
matter fields and analyzing the limit where these extra fields become very 
heavy. These new fields are usually taken to be pairs of chiral superfields 
transforming as $(N, {\bar N})$ under the color group. In the strong 
coupling regime, squark condensates will thefore form. These models, for the 
case of ($N-1$) flavours, were considered in 
Refs~\cite{Smilga97,Smilgazoo,Smilga98,SmilgaSUN}, where the analysis of the 
vacuum structure led the authors to conclude that the existence of BPS 
saturated domain walls was restricted to values for the mass $m$ of the
squark fields below a certain critical one. This jeopardized the idea 
of recovering pure gluodynamics by taking the limit $m \rightarrow \infty$, 
which is precisely the issue we want to revise here.

In order to do that let us consider supersymmetric QCD with $SU(N_c)$ gauge 
group and $N_f$ couples of chiral superfields $(Q_i,\bar{Q_i})$ 
transforming as $N_c,\bar{N_c}$. Non-perturbative effects become relevant at 
the scale $\Lambda$, where condensates form. The gaugino and squark 
colorless condensates are described by the following composite fields
\br
S & = & \frac{3}{32 \pi^2} {\rm Tr} W^2 \;\;, \nonumber \\
\\ 
M^i_j & = & Q^i \bar{Q}_j \; \; \; i,j=1,2,...,N_f \;\;, \nonumber
\er
where $W^2$ is the composite chiral superfield whose lowest component is
$\lambda \lambda$. In this regime, the relevant degrees of freedom are 
described by a Wess-Zumino model, as shown in Ref.~\cite{Venez82}. Its 
effective Langrangian is given by
\be
{\cal L} = \frac{1}{4} \int d^4 \theta \; {\cal K} + \frac{1}{2} \left[ 
\int d^2 \theta \; {\cal W} + {\rm h.c.} \right] \;\;,
\ee
where ${\cal K}$ is the K\"ahler potential and ${\cal W}$ is the 
superpotential
\be
{\cal W} = \frac {2}{3} S
\ln \frac{S^{N_c-N_f} \det M}{ \Lambda^{3 N_c - N_f} e^{N_c-N_f}} -
\frac {1}{2} {\rm Tr} (m M) \;\; ,
\label{TVY1}
\ee
with $m^j_k$ the mass matrix for the matter superfields. 
We will work in the flavour basis where this matrix is diagonal, and will
use the notation $ m^j_k = \delta^j_{k} m_j$. We will 
analyze the simple case where the corresponding eigenvalues are real. 
This superpotential has $N_c$ extrema labeled by the different phases of 
the gaugino condensate. At the minimum we have the gaugino condensate 
fixed to
\be
S_*^{N_c} = \left( \frac {3}{4} \right)^{N_f} \det m \;\;,
\label{gluino}
\ee
where, and from now on, we set $\Lambda=1$.
The matter fields are aligned with respect to the former and given by
\be
M_{i*}^j =  \delta_i^j \frac {1}{m_i} \frac {4}{3}  S_* \;\;.
\label{matter}
\ee
Finally, the superpotential at the minimum is proportional to the 
gaugino condensate
\be
{\cal W}_* = - \frac {2}{3} N_c S_* \;\;.
\ee
We want to study domain wall configurations that interpolate between the 
different minima. Here a technical problem appears: the superpotential has 
several branches associated with its logarithmic piece~\cite{Kogan98}. In 
the pure SUSY gluodynamics limit described by Veneziano and 
Yankielowicz~\cite{Venez82} this is a severe problem, since any 
configuration going from one vacuum to another has to cross this branch. 
This is not necessarily the case when we include other fields, given that 
the variation in the phase of the gaugino condensate can be partially 
compensated by these new fields. In this case, this will be done by matter 
fields. 

Let $(S,M)_a$ be a particular vacuum. We can continuously deform it into 
another vacuum, $(S,M)_b$. For this path in the configuration space, we
define $\delta$, $w_i$ such that
\br
S|_b     & = & e^{ i \delta } S |_a\;, \nonumber \\
\label{transf} \\
M_i^i |_b & = & e^{ i (\delta + 2 \pi w_i )}  M_i^i |_a \;, \; 
i=1,..., N_f \; . \nonumber
\er
Since, as mentioned above, matter at the minimum has to be aligned with 
respect to the gaugino condensate, $w_i$ must be some integer numbers. 
On the other hand, one necessary condition to avoid crossing the 
logarithmic branch along this general path is
\be 
(N_c-N_f) \delta +  \sum _i (\delta + 2 \pi w_i) = 0 \;\; ,
\ee
and then $\delta =  2 \pi  \frac{k}{N_c}$, where $k$ is the integer given 
by $k =  - \sum_i w_i  $.  

If we assume that there is a BPS domain wall connecting these two
vacua from $z=-\infty$ to $z=+\infty$, it will be described by the following 
differential equations
\br
{\cal K}_{S{\bar S}} \partial_z {\bar S}     & = & e^{i\gamma}  
\frac{ \partial{{\cal W}} } {\partial S } \;\; , \nonumber \\
\label{BPSeqs} \\
{\cal K}_{M{\bar M}} \partial_z {\bar M_i^i} & = & e^{i\gamma}  
\frac{ \partial{{\cal W}} } {\partial M^i_i } \;\;, \nonumber
\er
where ${\cal K}_{\phi{\bar \phi }} = \frac{\delta^2 {\cal K}} 
{\delta \phi \delta {\bar \phi}}$ is the induced metric from the K\"ahler 
potential ${\cal K}$, and $\gamma$ is given by
\be
\gamma = -\frac{1}{2}(\delta + \pi) = - \frac{k \pi}{ N_c} -\frac{\pi}{2}
\;\;.
\ee
Let us analyze the simplest case where the masses $m_i$ are degenerate.
We will assume symmetric boundary conditions for the matter fields. To be 
more precise, we will consider $w_i = - 1$  (and therefore $k=N_f$) for 
the path drawn by the domain wall. We can then assume that all the
matter condensates have the same $z$ dependence. Then the configuration is 
described by four real functions 
\br
M_i^i (z) & = & |M_*|  \; \rho(z)  e^{i\alpha (z) }  \;\;, \nonumber \\
\label{conds} \\
S (z) & = & |S_*| \;  R(z) e^{i\beta (z) } \;\;.  \nonumber
\er
Notice that we have defined $\rho(z), R(z)$ in such a way that 
$\rho(\pm \infty) = R (\pm \infty) = 1 $. On the other hand, $\alpha$ varies 
from $0$ to $2\pi(N_f/N_c - 1)$ and $\beta$ from $0$ to  
$2 \pi N_f/N_c$. A consistent ansatz under reflection 
$z\rightarrow -z$ is given by: $\rho(z)=\rho(-z)$, $R(z)=R(-z)$, 
$\beta(z) = 2 \pi  \frac{N_f}{N_c}  - \beta(-z)$ and
$\alpha(z) =  2 \pi ( \frac{N_f}{N_c} -1)   - \alpha(-z)$.
Then, we have the following  boundary conditions at $z=0$
\be
\label{incond} \\
\alpha(0) =  \pi \left( \frac{N_f}{N_c} -1 \right), \; \;
\beta(0)  =  \pi \frac{N_f}{N_c} 
\;\;. 
\ee
Eqs~(\ref{BPSeqs}) imply the following BPS constraint
\be
Im \left[ e^{i \gamma} {\cal W} (S, M_i^i) \right] = const \;\;.
\ee
In particular, at $z=0$ we have
\br
& - & R_0 \left[ (N_c-N_f) (\ln R_0 -1) +  N_f \ln \rho_0 \right] - N_f \rho_0
 \nonumber \\
& = & N_c \cos \left( \pi \frac{N_f}{N_c} \right) \;\;,  \label{orig}
\er
where $R_0 = R(0)$ and $\rho_0=\rho(0)$.

The case with $N_f = N_c-1$ has already been considered for $SU(2)$
\cite{Smilga97,Smilgazoo}, $SU(3)$ \cite{Smilga98}, and generalized to 
arbitrary $SU(N)$ in \cite{SmilgaSUN,Smilgatalk}. Since $k=N_c-1$, the 
corresponding domain walls 
connect a minimun and its nearest neighbour. In these papers it was shown 
that these domain walls are BPS states only for squark masses lower than 
some critical value, $m_*$, that depends on $N_c$ and the K\"ahler 
potential. The existence of this bound is related to the presence of two 
different BPS domain wall solutions for small enough values of $m$, which 
became identical at the critical value.
\begin{figure}
\centerline{  
\psfig{figure=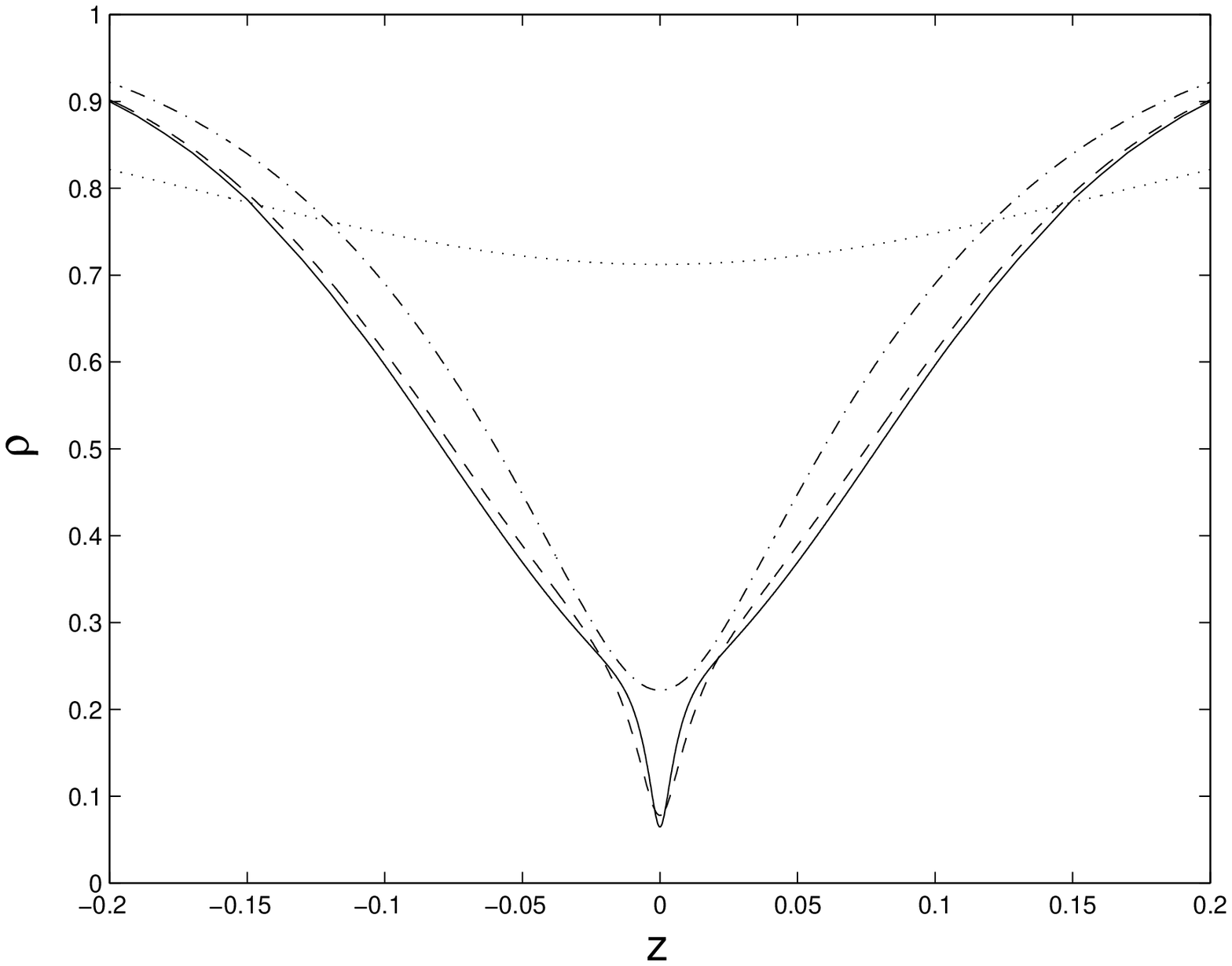,height=7cm,width=9cm,bbllx=0cm,bblly=7cm,bburx=21cm
,bbury=21cm}
}
\caption{}
{\footnotesize \noindent $\rho(z)$ as defined in Eqs~(\ref{conds}) 
versus $z$ (in units of ${\tilde{\Lambda}}^{-1}$), for $m = 2$ (dotted line), 
$20$ (dash-dotted), $100$ (dashed), $200$ (solid).}
\label{fig1}
\end{figure}

\begin{figure}
\centerline{  
\psfig{figure=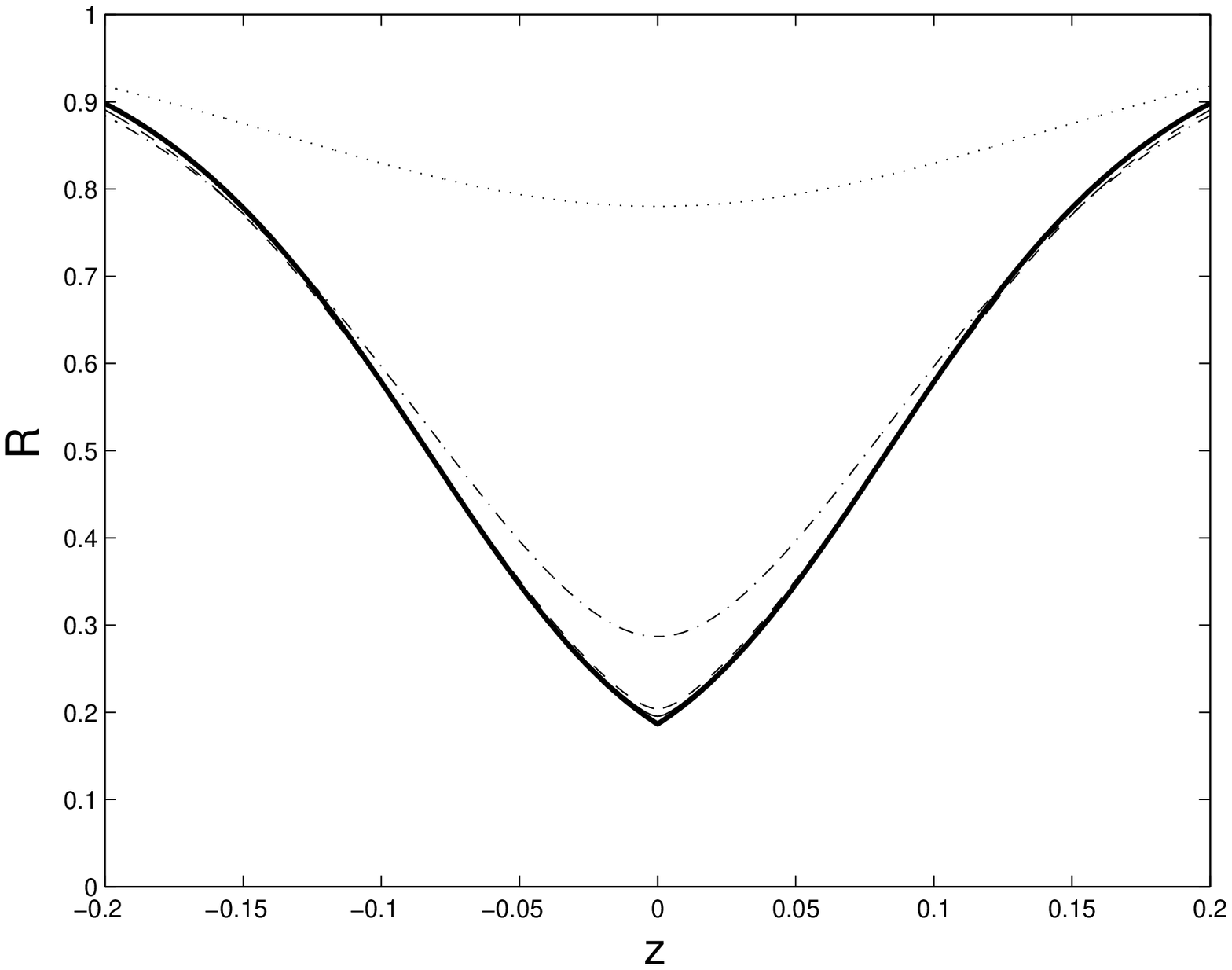,height=7cm,width=9cm,bbllx=0cm,bblly=7cm,bburx=21cm
,bbury=21cm}
}
\caption{}
{\footnotesize \noindent $R(z)$ as defined in Eqs~(\ref{conds}) versus $z$ 
(in units of ${\tilde{\Lambda}}^{-1}$), for $m = 2$ (dotted line), $20$ 
(dash-dotted), $100$ (dashed), $200$ (solid). The thick solid line corresponds
to the $m \rightarrow \infty$ solution given by Eq.~(\ref{limit}).}
\label{fig2}
\end{figure}

We have done a similar analysis for other values of $N_f$, 
using the same  K\"ahler potential , i.e. 
${\cal K} =  (S {\bar S})^{1/3}  + (M {\bar M})^{1/2} $.
Here we have worked in detail the case $N_c=3$, $N_f=1$, whereas other
cases will be presented elsewhere~\cite{us}. We have found that the 
equations can be solved for {\em all} values of the squark mass, and we have 
checked that the logarithmic branch is never crossed. The 
profiles for $\rho$ and $R$ are shown in Figs~\ref{fig1},\ref{fig2} for 
several values of $m$ (given in units of $\Lambda$), focusing on their 
central region. The spatial coordinate $z$ is expressed in units of 
${\tilde \Lambda}^{-1}$,  where ${\tilde \Lambda} = \Lambda  
(\frac{3 m}{4\Lambda})^{N_f/3N_c} $ is the effective QCD scale that arises 
in the large $m$ limit.

In our case there is only one BPS solution for every value of $m$. 
This can be understood analyzing both the large and small $m$ limit:

$\bullet$ When $m << \Lambda $ and $K_{S \bar S}$ is non singular,
we can integrate out the gaugino condensate by imposing 
$\partial {\cal W}/\partial S = 0$. The corresponding Wess-Zumino model 
describing the matter condensate has a BPS state with the required 
boundary conditions for all values of $N_f<N_c$. 
In the case analyzed by Smilga and Veselov~\cite{Smilga98,SmilgaSUN}, 
there is yet another BPS solution that cannot be described by integrating 
$S$ out and that corresponds to $S \sim 0 $. The existence of this domain 
wall is probably related to the fact that the K\"ahler metric $K_{S \bar S}$ 
is singular at this value~\cite{Kaplunovsky99}. In this limit $S \sim 0$, 
$\rho_0$  can be derived from Eq.~(\ref{orig}), and it is given by
\be
\rho_0 = - \frac{N_c}{N_f} \cos \left( \pi \frac{N_f}{N_c} \right) \;\;.
\ee
In our case, since $N_f/N_c < 1/2$, the resulting value for $\rho_0$ is never 
positive and therefore we do not find a second branch of solutions to the BPS 
equations.

$\bullet$ Let us turn now to analyze large mass values, 
$m >> \Lambda $. From Fig.~\ref{fig2} we see that there is a well defined 
gaugino condensate profile in the $m \rightarrow \infty $ limit. In fact, 
if we assume that this limit exists, the following constraints should 
apply in the asymptotic regions
\br
\rho (z) e^{i\alpha(z)}  & = &   R(z) e^{i  \beta(z)} \; \;   
(z << 1 /m) \;\;, \nonumber   \\
\label{inf} \\
\rho (z) e^{i\alpha(z)}  & = &   R(z) e^{i (\beta(z) - 2 \pi)}  \; \;   
(z >> 1 /m) \;\;.  \nonumber
\er
%
\begin{figure}
\centerline{  
\psfig{figure=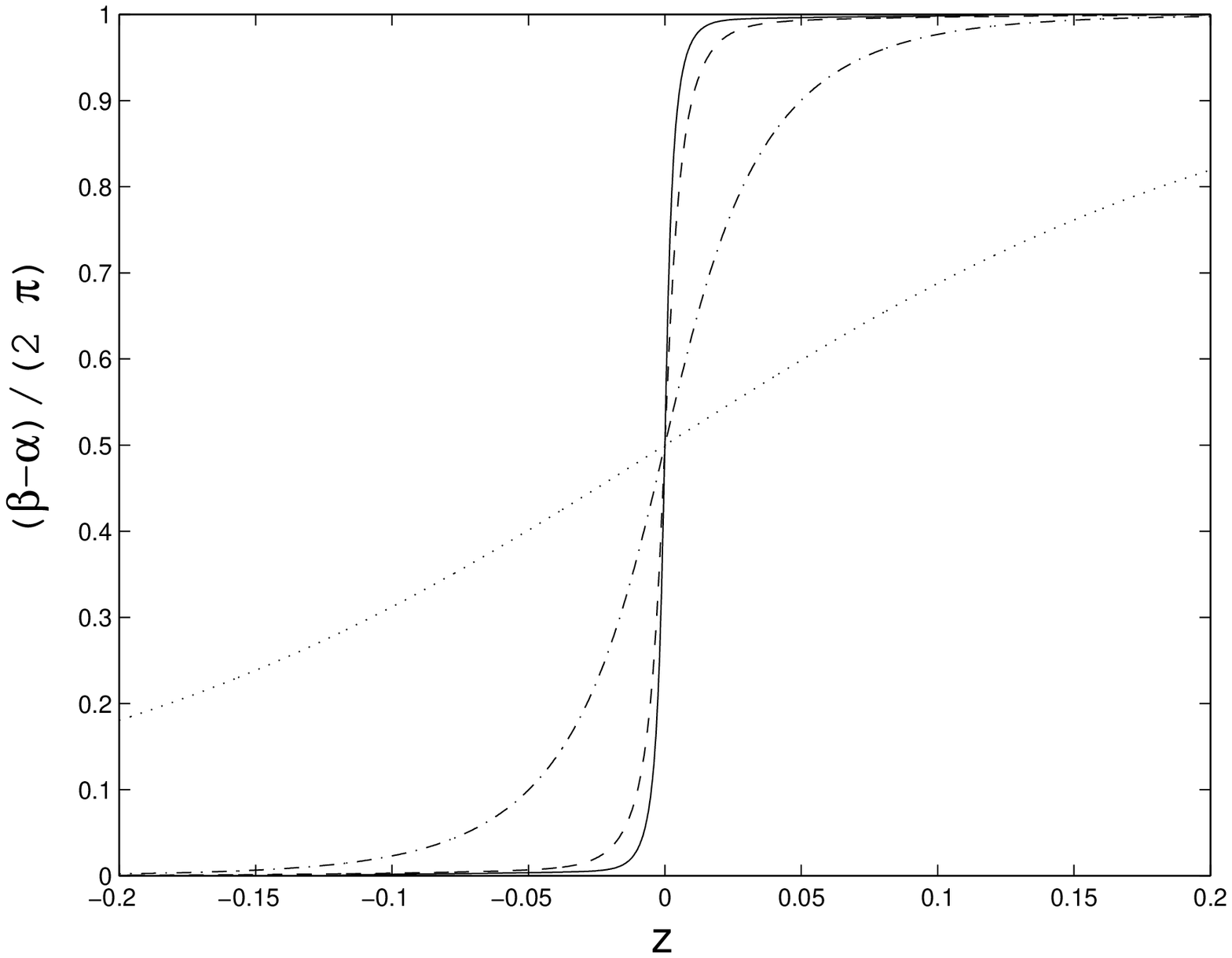,height=7cm,width=9cm,bbllx=0cm,bblly=7cm,bburx=21cm
,bbury=21cm}
}
\caption{}
{\footnotesize \noindent Plot of the combination $\frac{1}{2\pi} 
(\beta(z) - \alpha(z))$ versus $z$ (in units of ${\tilde{\Lambda}}^{-1}$), 
for $m = 2$ (dotted line), $20$ (dash-dotted), $100$ (dashed), $200$ (solid).}
\label{fig3}
\end{figure}
In Fig.~\ref{fig3} we have drawn the combination $\frac{1}{2\pi} 
(\beta(z) - \alpha(z))$, confirming the previous statement; also
a quick glance at Figs~1,2 tells us that both $\rho$ and $R$ follow
identical paths in the asymptotic regions.

Therefore, using Eq.~(\ref{inf}) in this large $m$ limit we can get rid of 
$\alpha$ and $\rho$ in the BPS equations. Also the BPS constraint involves 
only the gaugino condensate and can be written as
\be
Im \left\{ e^{i (\gamma + \tilde{\beta}(z))} R(z)  
\left[ \ln \left( R(z) e^{i \tilde{\beta}(z)} \right) -1 \right]
\right\} =  const \;\;,  
\label{split}
\ee
where $\tilde{\beta}(z) =  \beta(z)$ for $z<0$ and $\tilde{\beta}(z) =  
\beta(z) - 2\pi N_f/N_c$ for  $z>0$. This constraint allows us to 
express $\beta$ as a function of $R$, and we end up with the following 
BPS equation for $R(z)$
\br
\partial_z R (z) & = & 6 N_c (R(z))^{4/3} \tilde{\Lambda}  
\left\{ \cos (\gamma + \tilde{\beta} [R(z)]) \; \ln R(z)  \right. \nonumber \\
& - & \left. \sin (\gamma + \tilde{\beta} [R(z)]) \; \tilde{\beta} [R(z)] 
\right\} \;\;,
\label{limit}
\er
together with the boundary condition at the origin calculated from
Eq.~(\ref{split}), i.e.,
\be
R_0 (1-\ln R_0)  = \cos \left( \pi \frac{N_f}{N_c}  \right) \;\;.
\label{origin}
\ee
As we can see from Fig.~4, this equation has always a solution with 
$R_0 > 1$. There is also a solution with $R_0 < 1$ when $ 2 N_f < N_c $. 
We have seen that only the case $R_0 < 1$ gives a finite, domain wall like 
profile. We have also verified that the corresponding solution for $R(z)$ 
is precisely the large $m$ limit profile, which is represented by a thick
solid line in Fig.~2. 
Therefore the condition ensuring that there is just one branch of BPS
states at low $m$ values also guarantees the existence of the large 
$m$ limit case. 

\begin{figure}
\centerline{  
\psfig{figure=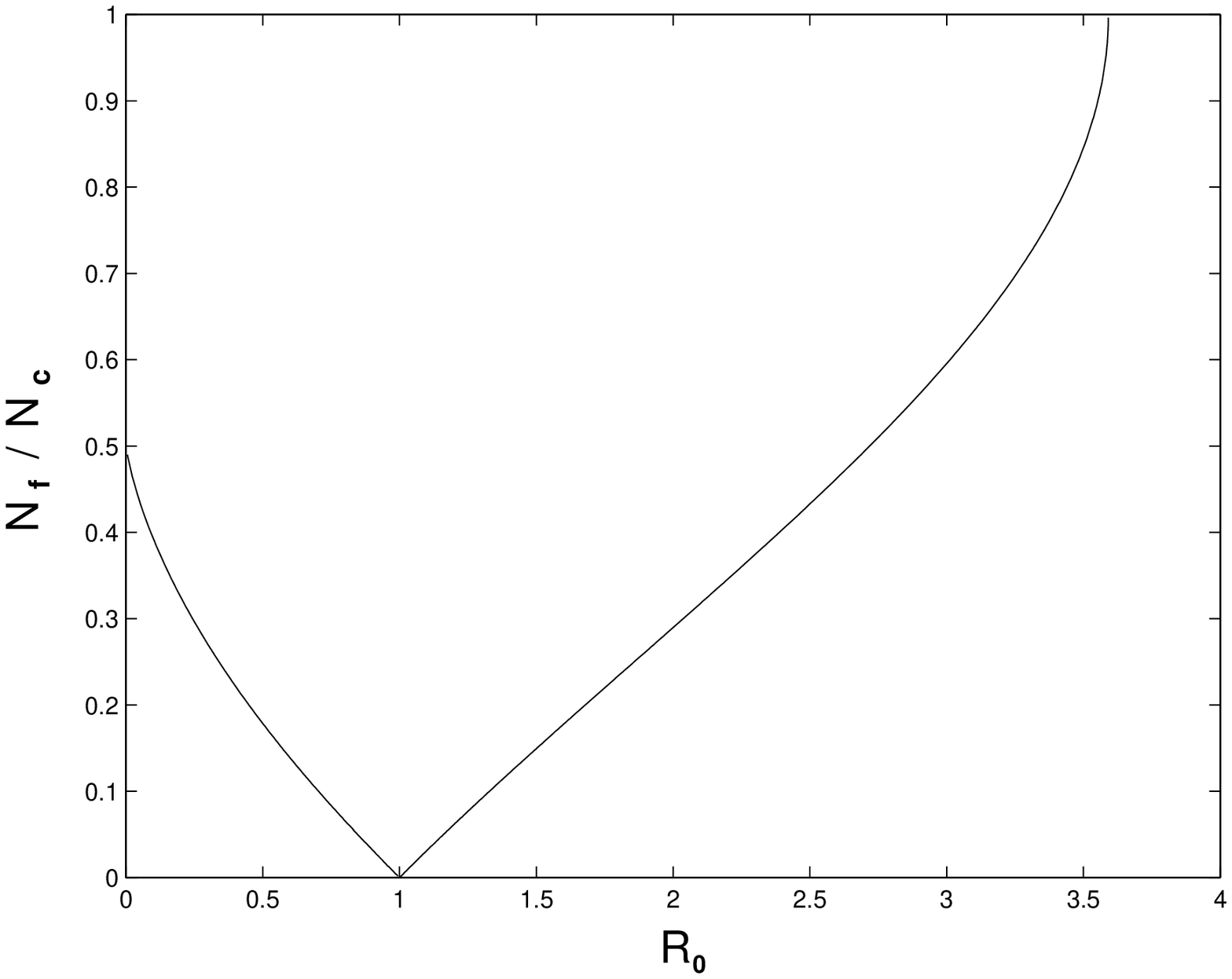,height=7cm,width=9cm,bbllx=0cm,bblly=7cm,bburx=21cm
,bbury=21cm}
}
\caption{}
{\footnotesize \noindent Contour plot of the constraint Eq.~(\ref{origin}) 
in the plane defined by the variables $R_0$ (x-axis) and $N_f/N_c$ (y-axis).}
\label{fig4}
\end{figure}

In summary, it is possible to build BPS domain walls in SQCD with 
$2 N_f <  N_c$, both in the weak coupling (Higgs) regime and in the strong 
coupling limit, where the theory approaches pure supersymmetric gluodynamics.
When the K\"ahler metric is non singular along the different configurations 
the existence of these solutions in the strong coupling regime can be 
understood by just analyzing the superpotential, as in the cases we have 
just shown.

\vspace{0.7cm}

We thank A.~Casas and M.~Hindmarsh for useful discussions, and
M.~Seco for his invaluable help with the computers.
The work of JMM was supported by CICYT of Spain (contract AEN98-0816).
JMM thanks the CERN Theory Division for hospitality, and we both
acknowledge the British Council/Acciones Integradas program for the 
financial support received through the grant HB1997-0073.

\widetext

\begin{thebibliography}{99}

\bibitem{KovnerS0}
A.~Kovner, M.~Shifman, \PRD{56}{97}{2396}, {\sf hep-th/9702174}.

\bibitem{Dvali97}
G.~Dvali, M.~Shifman, \PLB{396}{97}{64},{\em Erratum} \PLB{407}{97}{452}, 
{\sf hep-th/9612128}.

\bibitem{Kovner97}
A.~Kovner, M.~Shifman, A.V.~Smilga, \PRD{56}{97}{7978}, {\sf hep-th/9706089}.

\bibitem{Chibisov98}
B.~Chibisov, M.~Shifman, \PRD{56}{97}{7990}, {\em Erratum} 
\PRD{58}{98}{109901}, {\sf hep-th/9706141}.

\bibitem{Shifm98}
M.~Shifman, 6$^{th}$ International Symposium on Particles, 
Strings and Cosmology (PASCOS 98), Boston, MA, 22-27 Mar 1998, 
{\sf hep-th/9807166}.

\bibitem{Dvali99}
G.~Dvali, G.~Gabadadze, Z.~Kakushadze, {\sf hep-th/9901032}.

\bibitem{Witten97}
E.~Witten, \NPB{507}{97}{658}, {\sf hep-th/97106109}.

\bibitem{Smilga97}
A.V.~Smilga, A.I.~Veselov, \PRL{79}{97}{4529}, {\sf hep-th/9706217}.

\bibitem{Smilgazoo}
A.V.~Smilga, A.I.~Veselov, \NPB{515}{98}{163}, {\sf hep-th/9710123}.

\bibitem{Smilga98}
A.V.~Smilga, A.I.~Veselov, \PLB{428}{98}{303}, {\sf hep-th/9801142}.

\bibitem{SmilgaSUN}
A.V.~Smilga, \PRD{58}{98}{065005}, {\sf hep-th/9711032}.

\bibitem{Venez82}
G.~Veneziano, S.~Yankielowicz, \PLB{113}{82}{231}; 
T.~Taylor, G.~Veneziano, S.~Yankielowicz, \PLB{218}{83}{493}.

\bibitem{Kogan98}
I.I.~Kogan, A.~Kovner, M.~Shifman, \PRD{57}{98}{5195}, {\sf hep-th/9712046}.

\bibitem{Smilgatalk}
A.V.~Smilga, 3$^{rd}$ Workshop on Continuous Advances in QCD (QCD 98), 
Minneapolis, MN, 16-19 Apr 1998, {\sf hep-th/9807203}.

\bibitem{us}
B.~de~Carlos, M.~Hindmarsh, J.M.~Moreno, work in progress.

\bibitem{Kaplunovsky99}
V.S.~Kaplunovsky, J.~Sonnenschein, S.~Yankielowicz, {\sf hep-th/9811195}.

\end{thebibliography}
\end{document}